\documentclass[preprint,english,aps,superscriptaddress,prl]{revtex4-1}
\usepackage[T1]{fontenc}
\usepackage[latin9]{inputenc}
\usepackage{textcomp}
\usepackage{amsmath}
\usepackage{amsbsy}
\usepackage{graphicx}
\usepackage{amssymb}
\usepackage{esint}

\makeatletter

\providecommand{\LyX}{L\kern-.1667em\lower.25em\hbox{Y}\kern-.125emX\@}

\@ifundefined{textcolor}{}
{%
 \definecolor{BLACK}{gray}{0}
 \definecolor{WHITE}{gray}{1}
 \definecolor{RED}{rgb}{1,0,0}
 \definecolor{GREEN}{rgb}{0,1,0}
 \definecolor{BLUE}{rgb}{0,0,1}
 \definecolor{CYAN}{cmyk}{1,0,0,0}
 \definecolor{MAGENTA}{cmyk}{0,1,0,0}
 \definecolor{YELLOW}{cmyk}{0,0,1,0}
 }

\makeatother

\usepackage{babel}

\begin{document}

\preprint{}

\title{Spin Selection Rule-Based Sub-Millisecond Hyperpolarization of Nuclear Spins in Silicon}

\author{Felix Hoehne}
\email[corresponding author, email: ]{hoehne@wsi.tum.de}
\affiliation{Walter Schottky Institut, Technische Universit\"{a}t 
M\"{u}nchen, Am Coulombwall 4, 85748 Garching, Germany}

\author{Lukas Dreher}
\affiliation{Walter Schottky Institut, Technische Universit\"{a}t
M\"{u}nchen, Am Coulombwall 4, 85748 Garching, Germany}

\author{David P. Franke}
\affiliation{Walter Schottky Institut, Technische Universit\"{a}t
M\"{u}nchen, Am Coulombwall 4, 85748 Garching, Germany}

\author{Martin Stutzmann}
\affiliation{Walter Schottky Institut, Technische Universit\"{a}t
M\"{u}nchen, Am Coulombwall 4, 85748 Garching, Germany}

\author{Leonid S. Vlasenko}
\affiliation{A. F. Ioffe Physico-Technical Institute, Russian Academy of Sciences, 194021, St.~Petersburg, Russia}

\author{Kohei M. Itoh}
\affiliation{School of Fundamental Science and Technology, Keio University, 3-14-1 Hiyoshi, Kohuku-ku, Yokohama 223-8522, Japan}
\author{Martin S.~Brandt}
\affiliation{Walter Schottky Institut, Technische Universit\"{a}t
M\"{u}nchen, Am Coulombwall 4, 85748 Garching, Germany}
\begin{abstract}
In this work, we devise a fast and effective nuclear spin hyperpolarization scheme, which is in principle magnetic field and temperature independent. 
We use this scheme to experimentally demonstrate polarizations of up to 66$\%$ for phosphorus donor nuclear spins in bulk silicon, which are created within less than 100~$\mu$s in a magnetic field of 0.35~T at a temperature of 5~K. The polarization scheme is based on a spin-dependent recombination process via weakly-coupled spin pairs, for which the recombination time constant strongly depends on the relative orientation of the two spins. We further use this scheme to measure the nuclear spin relaxation time and find a value of $\sim$100~ms under illumination, in good agreement with the value calculated for nuclear spin flips induced by repeated ionization and deionization processes.
\end{abstract}
\maketitle
%

Nuclear spins in semiconductors have been intensively studied in the last decades as sensitive probes of the electronic structure of defects and, due to their exceptionally long decoherence times~\cite{Steger2012}, also as qubits for quantum information processing~\cite{Pla2013} or as a potential resource for a quantum memory~\cite{Morton2008}. However, their small magnetic moments and the resulting small polarization often impede their direct detection by nuclear magnetic resonance techniques, so that one has to resort to indirect detection schemes~\cite{Feher56ENDOR,dobers_electrical_1988,Smet04,stich1996}. An alternative strategy has focused on increasing the nuclear spin polarization above its thermal equilibrium value. Such hyperpolarization techniques have found widespread applications in magnetic resonance imaging~\cite{Moeller2002}, where in particular hyperpolarized silicon nanoparticles have been suggested as versatile agents for in-vivo imaging~\cite{Schroeder2006, Cassidy2013}. Further, in the context of spin-based quantum information processing, hyperpolarization schemes might be useful to initialize spin-based qubits~\cite{DiVincenzo2000,Simmons2011} or to improve the coherence times of electron spins coupled to a nuclear spin bath~\cite{Koppens2005}. 

Different hyperpolarization schemes of nuclear spins in silicon have been discussed, which mostly rely on the transfer of angular momentum from a polarized electron spin bath to the nuclear spins. While in direct semiconductors, circularly polarized light can be used to create spin-polarized electrons or holes~\cite{Puttisong2013}, this approach is not applicable to indirect semiconductors such as Si, where in most cases high magnetic fields and low temperatures are required~\cite{Vlasenko1986,Aptekar2009,Mccamey2009,Simmons2011,Lo2013}.
Recently, an efficient hyperpolarization procedure has been demonstrated for $^{31}$P in silicon based on the hyperfine selective optical excitation of donor-bound excitons, which however requires the use of ultrapure isotopically enriched $^{28}$Si~\cite{Yang2009}. In addition, all of these hyperpolarization schemes in silicon require time constants of at least 100~ms.

Here, we devise a fast and effective nuclear spin hyperpolarization scheme, based on a spin-dependent recombination process via weakly-coupled spin pairs~\cite{Kaplan78Spindep} as detailed below. We use this technique to experimentally demonstrate a large polarization of phosphorus donor nuclear spins in bulk silicon with natural isotope composition, which is created within less than 100~$\mu$s in a magnetic field of 0.35~T at a temperature of 5~K. 

\begin{figure}[!t]
\begin{centering}
\includegraphics[width=0.7\columnwidth]{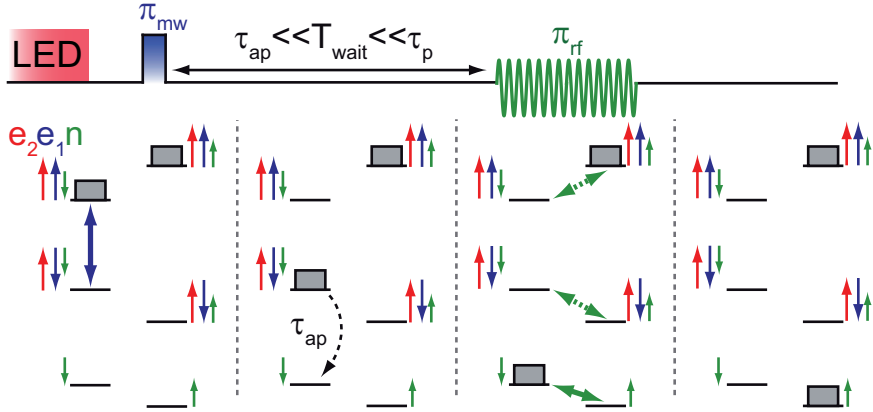}
\par\end{centering}
\caption{\label{fig:Figure1}
Pulse sequence for the hyperpolarization of nuclear spins with $I$=1/2 (green arrow, n) hyperfine-coupled to electron spins with $S$=1/2 (blue arrow, e$_1$). The electron spins form weakly coupled spin pairs with electron spins e$_2$ in spatial proximity (red arrow, e$_2$).
Only the four states with one orientation (spin up) of e$_2$ are shown here. A similar line of arguments can be applied to the four states with e$_2$ in the spin down state. The two states at the bottom denote the nuclear spin states of the $^{31}$P$^+$. See text for details.} 
\end{figure}

Considering a weakly coupled spin pair consisting of two electron spins e$_1$ and e$_2$ (red and blue arrows in Fig.~\ref{fig:Figure1}, resp.) with an additional nuclear spin n (green arrow) coupled by a hyperfine interaction to e$_1$, the difference in the recombination time constants $\tau_\mathrm{p}$ and $\tau_\mathrm{ap}$ of parallel and antiparallel electron spin pairs, resp., leads to large steady-state population differences under above-bandgap illumination~\cite{Hoehne_Timeconstants_2013}. States with both electron spins oriented in parallel are occupied (gray boxes) while antiparallel states are basically empty as shown exemplarily for e$_2$ spins up in Fig.~\ref{fig:Figure1}. This population difference can be transferred to the nuclear spins by the combination of a resonant microwave (mw) and radio-frequency (rf) $\pi$ pulse similar to a standard Davies ENDOR experiment~\cite{davies1974}, as illustrated in detail in the first three panels in Fig.~\ref{fig:Figure1}. However, since the recombination of antiparallel spin pairs takes place on timescales of the order of microseconds~\cite{Hoehne_Timeconstants_2013}, which is significantly shorter than the typical rf pulse length, this population transfer is rather inefficient~\cite{HoehneENDOR2011}. Therefore, by introducing a waiting period $T_\mathrm{wait}$ between the mw and rf pulse (Fig.~\ref{fig:Figure1}), which is chosen much longer than the recombination time of antiparallel spin pairs and much shorter than the recombination time of parallel spin pairs, all antiparallel spin pairs created by the mw $\pi$ pulse have recombined before the rf pulse. In addition, the illumination can be switched off during the pulse sequence to prevent new e$_1$-e$_2$ spin pairs to be formed by electron and hole capture processes~\cite{Dreher2012}. After these modifications, the population differences are stable on the much longer time scale $\tau_\mathrm{p}$, allowing for an efficient manipulation of the nuclear spins. 

This modified hyperpolarization scheme enables an almost complete transfer of the initial population difference between the antiparallel and parallel states to the nuclear spins by a single application of the pulse sequence shown in Fig.~\ref{fig:Figure1}. Since the initial population difference is determined by the parallel and antiparallel recombination rates and, therefore, is independent of the magnetic field, an almost complete polarization of the nuclear spins is possible even at low magnetic fields in contrast to most conventional hyperpolarization schemes, which transfer at most the thermal equilibrium electron spin polarization to the nuclear spins~\cite{Rosay2003}.

For $^{31}$P nuclear spins in silicon, at least two spin pairs can be employed for the presented hyperpolarization scheme, namely the $^{31}$P-P$_\mathrm{b0}$ spin pair at the Si/SiO$_2$ interface~\cite{Hoehne10} and the $^{31}$P-SL1 spin pair in $\gamma$-irradiated bulk silicon~\cite{Stich1995}.  
In the following, we will focus on the latter to experimentally demonstrate the hyperpolarization using a crystalline bulk phosphorus-doped Czochralski-grown silicon sample which has been exposed to $\gamma$-irradiation from a $^{60}$Co source. This creates oxygen-vacancy complexes which, under above-bandgap illumination, are excited into a metastable triplet state (SL1)~\cite{Brower1971} with a lifetime of the order of hundreds of microseconds at 5~K~\cite{Akhtar2012}. SL1 centers and $^{31}$P donors in spatial proximity form weakly coupled spin pairs giving rise to an efficient spin-dependent recombination process, which can be observed using electrically detected magnetic resonance (EDMR) as a resonant change in the photoconductivity~\cite{Stich1995}.  

\begin{figure}[!t]
\begin{centering}
\includegraphics[width=0.7\columnwidth]{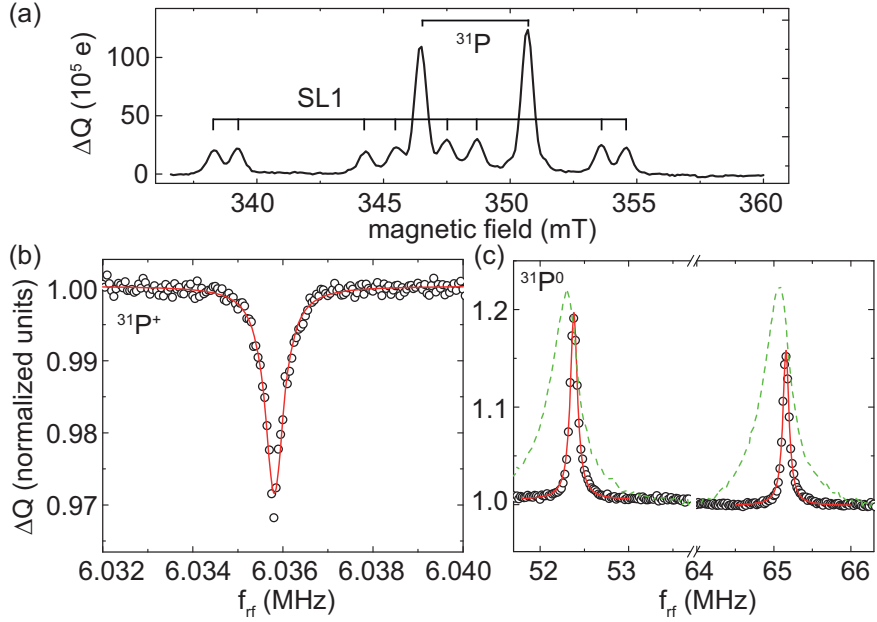}
\par\end{centering}
\caption{\label{fig:Figure1_b}
(a) EDMR spectroscopy of the $^{31}$P and SL1 electron spin transitions. 
Four additional peaks related to the SL1 are observed outside the magnetic field range shown here.
(b) and (c) Spectroscopy of the nuclear spin transitions of the ionized $^{31}$P$^+$ and the neutral $^{31}$P$^0$ (open circles). Resonance frequencies and peak widths are extracted from Lorentzian fits (red lines). For comparison, the spectroscopy of $^{31}$P$^0$ nuclear spins near the Si/SiO$_2$ interface is shown as well (dashed green line, data taken from~\cite{Dreher2012}).}
\end{figure}

To verify the presence of $^{31}$P-SL1 spin pairs in the sample, we first record a pulsed EDMR spectrum~\cite{Boehme03EDMR,Stegner06}.
To this end, we place the sample at 5~K in a dielectric resonator for pulsed ENDOR, illuminate it with above-bandgap light from an LED (wavelength 635~nm) and irradiate it with mw pulses of fixed length (70~ns) and frequency ($f_\mathrm{mw}$=9.739~GHz). The illumination intensity $I_\mathrm{LED}$ is calibrated by a photodetector inside the resonator. The current transients after the pulse sequence are amplified by a current amplifier, recorded with a fast data acquisition card and are box-car integrated, yielding a charge $\Delta Q$ which is proportional to the number of antiparallel spin pairs at the end of the mw pulse sequence~\cite{Boehme03EDMR}. Further details of the method are given in Ref.~\cite{Dreher2012}. 
The corresponding spectrum [Fig.~\ref{fig:Figure1_b}(a)] reveals the two hyperfine-split $^{31}$P peaks and eight peaks at magnetic field values in perfect agreement with the expected peak positions of the SL1 center~\cite{Brower1971}. The presence of an $^{31}$P-SL1 spin pair recombination process already indicated by the observation of both electron spin transitions in Fig.~\ref{fig:Figure1_b}(a) can be directly confirmed using electrically detected electron electron double resonance~\cite{Franke2013}.

To further assess the suitability of the $^{31}$P-SL1 spin pair for hyperpolarization, we determine the $^{31}$P-SL1 recombination time constants using a combination of pulsed optical excitation and pulsed spin manipulation~\cite{Dreher2012}. We find values of $\tau_\mathrm{ap}\approx$4~$\mu$s and $\tau_\mathrm{p}\approx$300~$\mu$s, confirming that antiparallel $^{31}$P-SL1 spin pairs recombine much faster than parallel spin pairs as required for the hyperpolarization scheme. We further characterize the spin transitions of the $^{31}$P nuclear spins both in the neutral and ionized state of the donor using pulsed electrically detected electron nuclear double resonance~\cite{HoehneENDOR2011,Dreher2012}. The spectra [Fig.~\ref{fig:Figure1_b}(b) and \ref{fig:Figure1_b}(c)] reveal a quenching of the echo signal at a frequency of $f_\mathrm{rf}$=6.0358(1)~MHz with an rf pulse excitation bandwidth-limited FWHM of 230~Hz, which corresponds to a nuclear $g$-factor of $g_\mathrm{n}$=-2.2606(3), in good agreement with the value of $g_\mathrm{n}$=-2.2601(3) observed at the Si/SiO$_2$ interface~\cite{Dreher2012}. Enhancements of the echo signal are found at frequencies of 52.38(1)~MHz and 65.15(1)~MHz (FWHM=100~kHz) corresponding to nuclear spin transitions of the neutral $^{31}$P donor.
The corresponding hyperfine interaction of $A$=117.54(2)~MHz is in good agreement with the value of $A$=117.523936(1)~MHz for $^{31}$P donors in bulk $^{28}$Si~\cite{Steger2011}. In contrast, for the $^{31}$P donors near the Si/SiO$_2$ interface [green dashed lines in Fig.~\ref{fig:Figure1_b}(c)], the nuclear spin transition frequencies correspond to a significantly smaller hyperfine constant of $A=117.31(2)$~MHz~\cite{Dreher2012},
which we attribute to strain at the surface~\cite{Wilson1961, Huebl2006Strain} caused by the evaporated metal contacts and their different thermal expansion coefficient compared to Si. Inhomogeneous strain might also explain the four times larger linewidth of these transitions. 


\begin{figure}[!t]
\centering
\includegraphics[width=0.7\columnwidth]{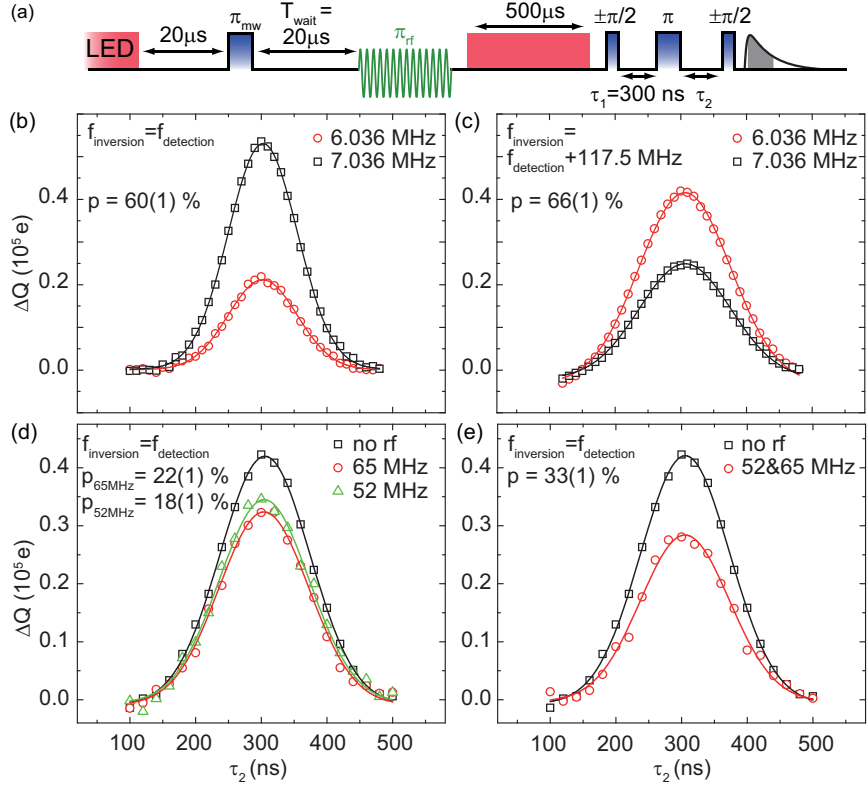}
\caption{
(a) Pulse sequence for the hyperpolarization of $^{31}$P nuclear spins. The resulting polarization is detected using a spin echo after new spin pairs have been generated by a 500~$\mu$s long LED pulse with an intensity of $I_\mathrm{LED}$=20~mW/cm$^2$. The time interval $\tau_\mathrm{ap}$=1.8~$\mu$s$\ll$ $T_\mathrm{wait}$=20~$\mu$s $\ll$ $\tau_\mathrm{p}$=260~$\mu$s between the mw inversion pulse and the rf pulse is chosen to ensure that all antiparallel spin pairs have recombined, while the time interval of 20~$\mu$s between switching off the LED and the first mw pulse is chosen much longer than the fall time of the LED pulse. (b) Detection spin echoes with a resonant ($f_\mathrm{rf}$=6.036~MHz) and an off-resonant rf pulse ($f_\mathrm{rf}$=7.036~MHz) with the mw inversion pulse and the detection echo resonantly exciting the high-field $^{31}$P hyperfine transition [cf. Fig.~\ref{fig:Figure1_b}(a)] resulting in a single shot nuclear spin polarization of $p$=60~\%. (c) Spin echo similar to (b), but with the detection echo on the high-field hyperfine transition and the inversion pulse on the low-field hyperfine transition resulting in $p$=66~\%. (d) Detection spin echoes with resonant rf pulses on the $^{31}$P$^0$ nuclear spin transitions (52.38~MHz and 65.15~MHz) and without rf pulse with polarizations of 18~\% and 20~\%, resp.. (e) Exciting both $^{31}$P$^0$ nuclear spin transitions, a polarization of 33~\% is achieved.}
\label{fig:Figure2}
\end{figure}

Based on the hyperpolarization scheme presented above, polarization of the $^{31}$P nuclear spins is created using the pulse sequence shown in Fig.~\ref{fig:Figure2}(a) (e$_1$=$^{31}$P, e$_2$=SL1) with an rf $\pi$ pulse on the nuclear spin transition of the $^{31}$P$^+$ at 6.036~MHz [cf. full green arrow in Fig.~\ref{fig:Figure1}(a)] or alternatively on one of the two $^{31}$P$^0$ nuclear spin transitions at 52.38~MHz or 65.15~MHz [cf. dotted green arrows in Fig.~\ref{fig:Figure1}(a)]. For the ideal case shown in Fig.~\ref{fig:Figure1}(a), a polarization of 100\% for the 6.036~MHz nuclear spin transition is expected after one application of the pulse sequence. In contrast, only 50\% can be achieved for the 52.38~MHz and 65.15~MHz transitions if only one of the two hyperfine-split $^{31}$P$^0$ nuclear spin transitions is excited~(Appendix A). Application of two subsequent rf $\pi$ pulses with 52.38~MHz and 65.15~MHz increases the maximum achievable polarization from 50\% to 100\% also for these transitions. 

The resulting nuclear spin polarization is determined after repopulating the donors by optical excitation for 500~$\mu$s to generate carriers in the conduction and valence bands and subsequent capture processes, assuming that the nuclear spin polarization is mostly unaffected by the repopulation process, which we will confirm below. 
Since only the nuclear spins of donors forming $^{31}$P-SL1 spin pairs are polarized, we use an electrically detected spin echo technique~\cite{Huebl08Echo,Hoehne2012} instead of conventional electron spin resonance to only measure the polarization of these nuclear spins.
The amplitude $\Delta Q_\mathrm{on}$ of the spin echo is compared with the spin echo amplitude $\Delta Q_\mathrm{off}$ after application of the same pulse sequence without or with off-resonant rf pulses. The measured nuclear spin polarization is given by $p=\left|1-\Delta Q_\mathrm{on}/\Delta Q_\mathrm{off}\right|$. To determine the value of $p$ obtained after a single repetition of the pulse sequence, we illuminate the sample for several hundreds of ms before applying the pulse sequence. This is much longer than the $^{31}$P nuclear spin relaxation time under illumination ($T_\mathrm{1n}$$\approx$100~ms) as determined below, leading to an effective randomization or reset of the nuclear spin system.  

Using the 6.036~MHz nuclear spin transition, we experimentally achieve a hyperpolarization of $\left|1-\Delta Q_\mathrm{on}/\Delta Q_\mathrm{off}\right|$=60~$\%$ for a single pulse sequence. Figure~\ref{fig:Figure2}(b) shows the corresponding spin echoes with a resonant rf pulse (black squares) and an off-resonant rf pulse (red circles) for $\tau_1$=300~ns as a function of $\tau_2$, with waiting times $\tau_1$ and $\tau_2$ after the first and second detection echo mw pulse, resp.. The values of $\Delta Q_\mathrm{on}$ and $\Delta Q_\mathrm{off}$ are determined by Gaussian fits (solid lines). The echo amplitude for the case of hyperpolarized nuclei is reduced compared with the reference as expected when the detection echo is measured on the same $^{31}$P electron spin hyperfine-split transition as the mw inversion pulse (cf.~Fig.~\ref{fig:Figure1}). Similarly, an increase of the echo amplitude is expected for the case that the detection echo and the inversion pulse are applied to different hyperfine transitions. To demonstrate this, we use a second mw source for the detection echo pulses detuned by the $^{31}$P hyperfine splitting of 117.5~MHz from the source for the inversion pulse. As shown in Fig.~\ref{fig:Figure2} (c), we indeed observe an increase of the echo amplitude for a resonant rf pulse corresponding to a hyperpolarization of 66~$\%$, also demonstrating that the observed polarization is not a spurious effect due to, e.g., heating by the strong rf pulses.

We can also use the 52.38~MHz and 65.15~MHz nuclear spin transition of the neutral donor for hyperpolarization, although we expect a smaller polarization value due to the lower fidelity of the rf $\pi$ pulse on the inhomogeneously broadened $^{31}$P$^0$ nuclear spin transition in Si with natural isotope composition. This is indeed observed as shown in Fig.~\ref{fig:Figure2}(d), where polarization values of 18$\%$ and 22$\%$ are achieved for the 52~MHz and 65~MHz nuclear spin transitions, resp.. The polarization can be increased to 33$\%$ by applying two subsequent rf pulses on both nuclear spin transitions as shown in Fig.~\ref{fig:Figure2}(e).
     
The nuclear spin hyperpolarization values of 60~$\%$ and 66~$\%$ exceed the thermal equilibrium polarization of the $^{31}$P nuclear spins at 0.35~T and 5~K of 3$\cdot 10^{-5}$ by a factor of $\approx 2 \cdot 10^4$ and even exceed the thermal equilibrium electron spin polarization of $\approx$5\% under these conditions by a factor of 12. This is achieved after a single repetition of the pulse sequence taking less than 100~$\mu$s, demonstrating that we have realized a fast and efficient nuclear spin hyperpolarization scheme.

The fidelity of the polarization scheme depends on several aspects. First, the excitation bandwidth of the mw and rf polarization pulses has to be much larger than the linewidth of the electron and nuclear spin transitions to allow for high-fidelity $\pi$ pulses. For both, the $^{31}$P electron and the $^{31}$P$^+$ nuclear spin, the excitation bandwidths of $\approx$50~MHz and $\approx$20~kHz are much larger than the linewidths of $\approx$8~MHz~\cite{Lu2011} and 230~Hz, resp.. From these values, we estimate a pulse fidelity of \textgreater 90\% and $\approx$100\% for the mw and rf $\pi$ pulse, resp.. Further, the difference between $\tau_\mathrm{ap}$ and $\tau_\mathrm{p}$ has to be sufficiently large so that the condition $\tau_\mathrm{ap} \ll T_\mathrm{wait}+T_\mathrm{rf} \ll \tau_\mathrm{p}$ can be fullfilled, where we also take the length $T_\mathrm{rf}$ of the rf pulse into account. For the $^{31}$P-SL1 spin pair, we estimate that in addition to all antiparallel spin pairs, also a fraction of 1-$\exp(-(T_\mathrm{wait}+T_\mathrm{rf})/\tau_\mathrm{p}))$$\approx$0.2 of parallel spin pairs recombines until the end of the rf pulse. Although only a rough estimate, this partly explains the observed maximum nuclear spin polarization of $\approx$66\%. A more detailed analysis should include a detailed model of the spin pair dynamics~\cite{Hoehne_Timeconstants_2013} and also take into account the variation of recombination time constants over the spin pair ensemble.

\begin{figure}[!t]
\centering
\includegraphics[width=0.7\columnwidth]{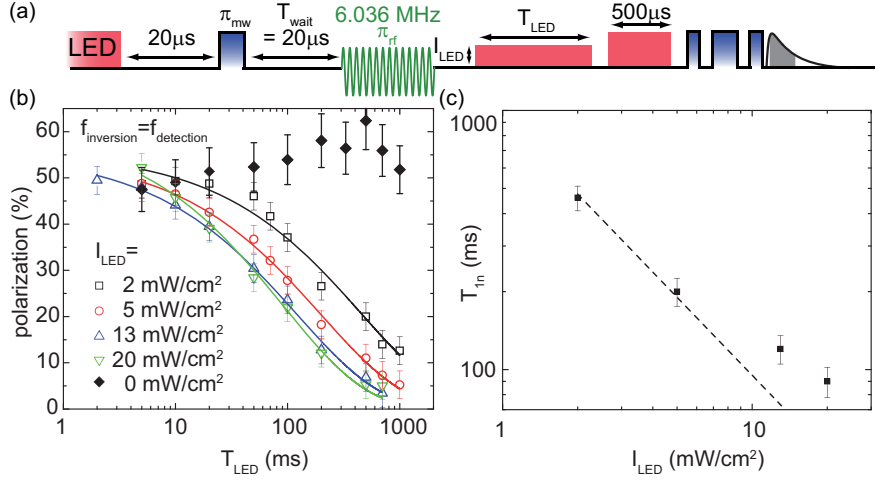}
\caption{
(a) Pulse sequence to measure the $^{31}$P nuclear spin relaxation time $T_\mathrm{1n}$ for different illumination intensities I$_\mathrm{LED}$.
(b) Polarization as a function of the optical excitation pulse length $T_\mathrm{LED}$ for different illumination intensities on a log-log scale (open symbols). For comparison, the data without a light pulse is shown as well (full diamonds). The nuclear spin relaxation time $T_\mathrm{1n}$ as shown in (c) is determined by single exponential fits (solid lines) of the data in (b). The $I_\mathrm{LED}^{-1}$ dependence (dashed line) is a guide for the eye.}
\label{fig:Figure3}
\end{figure} 

Having established a large single shot hyperpolarization, we proceed by measuring the nuclear spin relaxation time $T_{\mathrm{1n}}$ for different illumination intensities. To this end, we use the hyperpolarization and detection pulse sequence discussed above and apply an additional light pulse of variable length $T_\mathrm{LED}$ and intensity $I_\mathrm{LED}$ between the rf $\pi$ pulse and the detection [see Fig.~\ref{fig:Figure3}(a)]. Again, a more than 500~ms long illumination pulse is applied before each repetition of the pulse sequence to ensure that the nuclear spins are randomized. Figure~\ref{fig:Figure3}(b) shows the decay of the nuclear spin polarization as a function of $T_\mathrm{LED}$ for different $I_\mathrm{LED}$ (open symbols) measured for the $^{31}$P$^+$ nuclear spin transition as in Fig.~\ref{fig:Figure2}(b). We observe a nuclear spin relaxation time of $T_\mathrm{1n}\approx$100~ms for the highest illumination intensity as determined by a single exponential fit. 
For low $I_\mathrm{LED}$, $T_\mathrm{1n}$ decreases approx. $\propto I_\mathrm{LED}^{-1}$ as indicated by the dashed line in Fig.~\ref{fig:Figure3}(b).
Without illumination, a small increase rather than a decrease of the polarization is observed for time intervals as long as 1~s [full diamonds in Fig.~\ref{fig:Figure3}(b)].    
The latter observation is in line with the very long $^{31}$P nuclear spin relaxation time of $\sim$10 hours that has been found in bulk Si:P samples at 0.32~T and 1.25~K without above-bandgap illumination~\cite{FeherII59Relaxation}.

$T_{\mathrm{1n}}$ is shortened by optical excitation of carriers into the conduction and valence bands. 
Possible relaxation mechanisms are, e.g., the scattering of conduction band electrons with the $^{31}$P nuclei, leading to spin flip-flop processes, which however predicts relaxation times of several hours at $B_0$$\approx$0.3~T~\cite{Overhauser1953,FeherII59Relaxation,Mccamey2009}. 
 Nuclear spin flips can also be induced by repeated ionization and deionization of the $^{31}$P donor because of the mixing of the high-field eigenstates by the hyperfine interaction~\cite{Pla2013}. 

 
 The probability of a nuclear spin flip for each ionization/deionization process is given by $P_\mathrm{flip}=\sin(\eta/2)^2\approx3.6\cdot10^{-5}$~\cite{Pla2013}, where $\eta=\arctan(A/f_\mathrm{P})$ denotes the mixing angle as defined, e.g., in Ref.~\cite{Steger2011}, with the $^{31}$P hyperfine coupling $A$=117.5~MHz and the $^{31}$P electron spin Larmor frequency $f_\mathrm{P}$=9.798~GHz at $B_0$=350.3~mT. A detailed analysis of the time evolution of the spin system~(Appendix B) shows that for high $I_\mathrm{LED}$ $T_\mathrm{1n}=\tau_\mathrm{ap}/(P_\mathrm{flip})$ with $\tau_\mathrm{ap}$=4~$\mu$s results in $T_\mathrm{1n}$=110~ms, in very good agreement with the experimentally observed relaxation time. For lower $I_\mathrm{LED}$, the formation rate of new spin pairs by electron and hole capture processes decreases $\propto I_\mathrm{LED}^{-1}$~\cite{Hoehne_Timeconstants_2013}, resulting in an increase of the average time the spin pair spends in the ionized state. This reduces the ionization rate and therefore the nuclear spin flip rate, explaining the observed increase of $T_\mathrm{1n}$ with decreasing $I_\mathrm{LED}$ [cf.~Fig.~\ref{fig:Figure3}(c)].   
 
%

To summarize, we have demonstrated a fast and effective nuclear spin polarization scheme for $^{31}$P nuclear spins in natural Si at 5~K achieving a polarization of 66\% within less than 100~$\mu$s. The polarization scheme does not rely on thermal equilibrium spin polarizations and therefore works at easily accessible magnetic fields and temperatures. 
We further note, that no electrical contacts are needed to create the polarization, they were used solely for the measurement of the polarization.  
The density of polarized nuclear spins in the studied sample is at most $\sim$10$^{12}$~cm$^{-3}$, limited by the concentration of SL1 centers, and therefore orders of magnitude too small for a possible application in magnetic resonance imaging.
However, systems with a much larger density of spin pairs can be envisaged like, e.g., P-doped silicon nanoparticles. Here, the nanoparticle diameter, the doping concentration and the density of dangling bond defects can be adjusted such that each nanoparticle contains one P donor and one defect with high probability~\cite{almeida2012,Niesar2012}, so that spin pair densities of more than 10$^{17}$~cm$^{-3}$ could be achieved, sufficient for nuclear magnetic resonance imaging. For such $^{31}$P-P$_\mathrm{b0}$ spin pairs, we have obtained a nuclear spin polarization of $\approx$ 30\% at the Si:P/SiO$_2$ interface, indicating that the described method is also applicable to them.   
      
The work was funded by DFG (Grant No. SFB 631, C3 and Grant No. SPP 1601, Br 1585/8), the JST-DFG Strategic Cooperative Program on Nanoelectronics, the Core-to-Core Program by JSPS, FIRST, and a Grant-in-Aid for Scientific Research and Project for Developing Innovation Systems by MEXT.

\appendix
\section{Polarization Scheme of $^{31}$P$^0$ Nuclear Spins}
\begin{figure}[t]
\begin{centering}
\includegraphics[width=0.7\columnwidth]{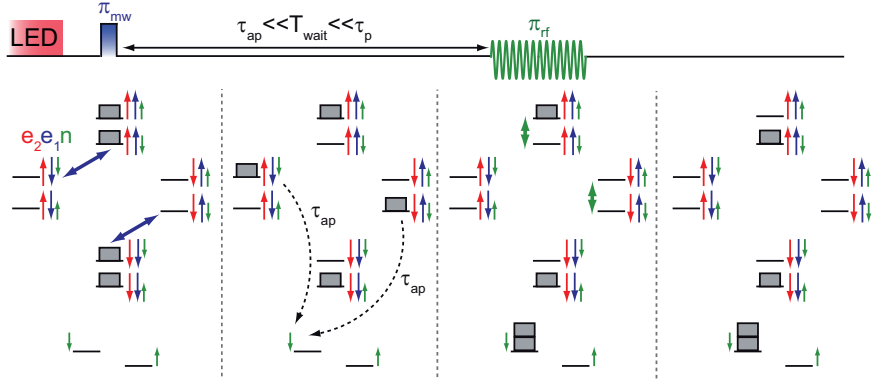}
\par\end{centering}
\caption{\label{fig:Fig1_SOM}
Hyperpolarization pulse sequence with an extended level scheme showing all eight states of the spin systems. For an rf $\pi$ pulse on one of the two $^{31}$P$^0$ nuclear spin transitions, a maximum final polarization of 50\% can be achieved.}
\end{figure}

In Fig.~\ref{fig:Figure1}, we sketched the nuclear spin hyperpolarization scheme showing only four of the eight states of the spin system. We argued, that for an rf pulse on the $^{31}$P$^+$ nuclear spin transition a maximum polarization of 100\% is expected, while for rf pulses on the two $^{31}$P$^0$ nuclear spin transitions this value is reduced to 50\%. The reason for the latter can most easily be seen by visualizing the spin state populations for all eight states of the spin system as shown in Fig.~\ref{fig:Fig1_SOM}. There, we exemplarily sketch the hyperpolarization sequence for an rf $\pi$ pulse resonant with one of the $^{31}$P$^0$ nuclear spin transitions (green arrow) resulting in a population of 1/4 for the nuclear spin up states and 3/4 for the nuclear spin down states, which corresponds to a polarization of 50\%. In contrast, an rf $\pi$ pulse on the $^{31}$P$^+$ transition or two rf $\pi$ pulses on both $^{31}$P$^0$ nuclear spin transitions completely polarize the nuclear spins.

\section{Nuclear Spin Relaxation under Illumination}

In this section, we discuss the relaxation time of the $^{31}$P nuclear spins caused by the ionization/deionization process under above-bandgap illumination.
The hyperfine interaction leads to a mixing of states with different nuclear spin orientations. Upon ionization, the mixed state is projected on the $^{31}$P$^+$ nuclear spin up or spin down eigenstate with a certain probability $P_\mathrm{flip}$ of a nuclear spin flip, which we calculate below. The nuclear spin relaxation rate $1/T_\mathrm{1n}$ is then calculated by multiplying $P_\mathrm{flip}$ with the effective recombination rate.

For the $^{31}$P electron spin e$_1$ coupled to the $^{31}$P nuclear spin n by a hyperfine interaction $A$, the Hamiltonian is given by
\begin{equation}
\frac{\hat{H}}{h}=f_\mathrm{P}\cdot S_\mathrm{z}+A\cdot \vec{S}\cdot\vec{I},
\end{equation} 
with the $^{31}$P electron spin Larmor frequency $f_\mathrm{P}$, and the electron spin and nuclear spin operators $\vec{S}$ and $\vec{I}$, respectively.
For $f_\mathrm{P}\gg A$, the eigenstates of $\hat{H}$ are given by $\left|\uparrow\Uparrow\right\rangle$,$\left|\uparrow\Downarrow\right\rangle$, $\left|\downarrow\Uparrow\right\rangle$, and $\left|\downarrow\Downarrow\right\rangle$, where the first arrow denotes the electron spin state and the second arrow the nuclear spin state. The hyperfine coupling leads to a mixing of these high-field states and the eigenstates of $\hat{H}$ become
\begin{equation}
\begin{split}
\left|1\right\rangle & = \left|\uparrow\uparrow\Uparrow\right\rangle \\
\left|2\right\rangle & = \cos(\frac{\eta}{2})\left|\uparrow\uparrow\Downarrow\right\rangle+\sin(\frac{\eta}{2})\left|\uparrow\downarrow\Uparrow\right\rangle \\
\left|3\right\rangle & = \sin(\frac{\eta}{2})\left|\uparrow\uparrow\Downarrow\right\rangle+\cos(\frac{\eta}{2})\left|\uparrow\downarrow\Uparrow\right\rangle \\
\left|4\right\rangle & = \left|\uparrow\downarrow\Downarrow\right\rangle ,
\end{split}
\end{equation}
where we have also introduced the electron spin e$_2$ denoted by the very first arrow as the recombination partner of the $^{31}$P electron spin shown exemplarily only for spin up.
The mixing angle $\eta$ is defined as 
\begin{equation}
\tan(\eta)=\frac{A}{f_\mathrm{P}}.
\end{equation}
For the experimental conditions in this work ($f_\mathrm{P}\approx$9.798~GHz, $A$=117.5~MHz), we obtain $\sin(\eta/2)$=6$\cdot 10^{-3}$, so that the nuclear spin flip probability is $P_\mathrm{flip}=\sin(\eta/2)^2=4\cdot 10^{-5}$.


We further introduce an operator $\hat{T}$ describing the spin-dependent transition into the ionized $^{31}$P$^+$ state denoted by $\left|f_\Uparrow\right\rangle$ ($\left|f_\Downarrow\right\rangle$) for the $^{31}$P nuclear spin up (down). We assume that $\hat{T}$ conserves the nuclear spin state. Since the recombination time constants of the unmixed states $\left|1\right\rangle$ and $\left|4\right\rangle$ are $\tau_\mathrm{p}$ and $\tau_\mathrm{ap}$, resp., the matrix elements of $\hat{T}$ 
are given by 
\begin{equation}
\begin{split}
\left|\left\langle 1\right|\hat{T}\left|f_\Uparrow\right\rangle\right|^2 = \left|\left\langle \uparrow\uparrow\Uparrow\right|\hat{T}\left|f_\Uparrow\right\rangle\right|^2 & \propto 1/\tau_\mathrm{p}\\
\left|\left\langle 1\right|\hat{T}\left|f_\Downarrow\right\rangle\right|^2 = \left|\left\langle \uparrow\uparrow\Uparrow\right|\hat{T}\left|f_\Downarrow\right\rangle\right|^2 & = 0\\
 \left|\left\langle 4\right|\hat{T}\left|f_\Uparrow\right\rangle\right|^2 = \left|\left\langle \uparrow\downarrow\Downarrow\right|\hat{T}\left|f_\Uparrow\right\rangle\right|^2 & = 0\\
\left|\left\langle 4\right|\hat{T}\left|f_\Downarrow\right\rangle\right|^2 = \left|\left\langle \uparrow\downarrow\Downarrow\right|\hat{T}\left|f_\Downarrow\right\rangle\right|^2  & \propto 1/\tau_\mathrm{ap}.
\end{split}
\end{equation}

A nuclear spin flip occurs, when one of the two mixed states $\left|2\right\rangle$ and $\left|3\right\rangle$, which are mainly nuclear spin down and up, respectively, is projected on the $^{31}$P$^+$ state with the opposite nuclear spin orientation.
The corresponding matrix elements are given by
\begin{equation}
\begin{split}
\left|\left\langle 2\right|\hat{T}\left|f_\Uparrow\right\rangle\right|^2 & = \\
\left|\cos(\frac{\eta}{2})\left\langle \uparrow\uparrow\Downarrow \right|\hat{T}\left|f_\Uparrow\right\rangle +
\sin(\frac{\eta}{2})\left\langle \uparrow\downarrow\Uparrow \right|\hat{T}\left|f_\Uparrow\right\rangle\right|^2 & \propto \sin(\frac{\eta}{2})^2 /\tau_\mathrm{ap} \end{split}
\end{equation}
and
\begin{equation}
\begin{split}
\left|\left\langle 3\right|\hat{T}\left|f_\Downarrow\right\rangle\right|^2 & = \\
\left|\cos(\frac{\eta}{2})\left\langle \uparrow\downarrow\Uparrow \right|\hat{T}\left|f_\Downarrow\right\rangle +
\sin(\frac{\eta}{2})\left\langle \uparrow\uparrow\Downarrow \right|\hat{T}\left|f_\Downarrow\right\rangle\right|^2 & \propto \sin(\frac{\eta}{2})^2 /\tau_\mathrm{p}. \end{split}
\end{equation} 
Since $\tau_\mathrm{ap} \ll \tau_\mathrm{p}$, we can neglect the latter, so that the nuclear spin relaxation rate is given by
\begin{equation}
\frac{1}{T_\mathrm{1n}}= \frac{\sin(\eta/2)^2}{\tau_\mathrm{ap}},
\label{eq:}
\end{equation}
which is just the nuclear spin flip probability $P_\mathrm{flip}=\sin(\eta/2)^2$ multiplied with the effective recombination rate $1/\tau_\mathrm{ap}$.
For $\sin(\eta/2)$=6$\cdot 10^{-3}$ and $\tau_\mathrm{ap}$=4~$\mu$s, we obtain $T_\mathrm{1n}$=110~ms in very good agreement with the experimentally observed value. 

In Fig.~\ref{fig:Figure3}(b), we observe that $T_\mathrm{1n}$ becomes significantly longer for decreasing illumination intensity $I_\mathrm{LED}$. This can be understood by taking into account that to complete an ionization/deionization cycle, new spin pairs have to be generated after the recombination process. Since this involves the capture of an electron from the conduction band by the $^{31}$P$^+$, the characteristic time constant of the spin pair generation $\tau_\mathrm{g}$ depends on the density of conduction electrons and, therefore, is $\propto I_\mathrm{LED}^{-1}$~\cite{Hoehne_Timeconstants_2013}. For high $I_\mathrm{LED}$, $\tau_\mathrm{g}\ll \tau_\mathrm{p}$, so that almost instantaneously after a recombination process, new spin pairs are formed and, therefore, $T_\mathrm{1n}$ is determined as discussed above. However, for lower $I_\mathrm{LED}$, $\tau_\mathrm{g}$ eventually becomes larger than $\tau_\mathrm{p}$, so that after a recombination process, the system spends some time in the ionized state. This reduces the ionization/deionization rate thereby increasing $T_\mathrm{1n}$ as observed in Fig.~\ref{fig:Figure3}(b). Note, that the life time of state $\left|2\right\rangle$ is $\sim\tau_\mathrm{p}$ rather than $\sim\tau_\mathrm{ap}$ since $\cos(\eta/2)^2/\tau_\mathrm{p}\gg\sin(\eta/2)^2/\tau_\mathrm{ap}$ and, therefore, $T_\mathrm{1n}$ starts increasing already for $\tau_\mathrm{g}$\textless$\tau_\mathrm{p}$.

\end{document}